\documentclass[12pt]{article}
\usepackage{graphicx}
\usepackage{amsmath}
\usepackage{amsfonts}
\usepackage{amssymb}

\begin{document}
% \title{  }

\begin{center}

\Large\bf{Simulation of boron diffusion during low-temperature
annealing of implanted silicon}
\\[2ex]

\normalsize
\end{center}

\begin{center}
\textbf{O. I. Velichko and A. P. Kavaliova}
\end{center}

Department of Physics, Belarusian State University of Informatics

and Radioelectronics, 6, P.~Brovki Street, Minsk, 220013 Belarus

{\it E-mail addresses: velichkomail@gmail.com} (Oleg Velichko)

\begin{abstract}
Modeling of ion-implanted boron redistribution in silicon crystals
during low-temperature annealing with a small thermal budget has
been carried out. It was shown that formation of ``tails'' in the
low-concentration region of impurity profiles occurs due to the
long-range migration of boron interstitials.
\end{abstract}

% begin{keyword}
\textit{\textbf{Keywords}}: Diffusion; Annealing; Doping Effects;
Boron; Silicon
% end{keyword}\documentclass [12pt]{article}

\textbf{PACS} 66.30.-h, 61.72.Uf, 85.40.Ry

  \section{Introduction}

It is well known that device scaling requires decreasing all
vertical and lateral dimensions of transistors At present, a
low-energy high-dose ion implantation is used for the formation of
shallow junctions However, we need to undergo silicon substrate to
subsequent annealing to avoid defects created by ion implantation.
During thermal treatment of ion implanted layers a great amount of
nonequilibrium self-interstitials is generated, which results in
transient enhanced diffusion (TED) of impurity atoms. Intensive
TED prevents the formation of shallow junctions especially for
boron doped regions (see, for example,
\cite{Stolk-97,Jain-02,Shao-03,Philippe-11}). Therefore, the
problem of formation of doped regions with abrupt impurity
distributions in vertical and lateral directions requires
annealing imposed with an extremely small thermal budget for the
TED to be decreased \cite{ITRS-11}. One of the techniques to
provide a small thermal budget and solve this problem is the use
of low-temperature annealing of short duration
\cite{Napolitani-99}. Another technique widely used to suppress
transient enhanced diffusion of ion-implanted impurity is the
introduction of boron into the silicon layer preamorphized by
implantation of Ge ions
\cite{Cristiano-04,Hamilton-07,Yeong-08,Salvador-10}. As a result
of the solid phase epitaxial recrystallization (SPER) of the
amorphous layer at the initial stage of annealing a perfect
crystalline structure without defects visible by means of
electronic microscopy is formed in the array of the implanted
boron. Nevertheless, in this case TED is also observed during the
subsequent annealing, but of absolutely another character, and
characterized by a smaller intensity. Indeed, at low annealing
temperatures an extended ``tail'' in the low-concentration region
of boron profile is formed in all the cases mentioned above. This
``tail'' represents a direct line for logarithmic scale on the
concentration axis.

It was shown in \cite{Velichko-10,Velichko-VITT-11,Velichko-12}
that the majority of the cases of ``tail'' formation in the
low-concentration region of ion-implanted impurity distribution,
including boron implantation in preamorphized silicon, is related
to the phenomenon of a long-range migration of nonequilibrium
impurity interstitials, and we can neglect channeling of a
fraction of implanted atoms if ion implantation of In, P, and Ga
is performed in a random direction or B is implanted in a
preamorphized layer. In \cite{Kniazhava-10,Velichko-2012}, a
similar theoretical investigation and simulation of impurity
redistribution were carried out for boron implantation in a random
direction in the crystalline silicon. It has been shown that the
long-range migration of nonequilibrium boron interstitials is
responsible for the ``tail'' formation in a crystal lattice
damaged by boron ions implanted at medium fluencies which are
insufficient to create an amorphous layer. For example, it was
found in \cite{Velichko-2012}  that the average migration length
of boron interstitials $l_{AI}$ is equal to 24 nm for annealing at
a temperature of 600 $^{\circ}$C for 10 s. This value is two times
greater than the average migration length of boron interstitials
$l_{AI}$ = 12 nm and $l_{AI}$ = 11 nm found for boron diffusion in
the preamorphized and recrystallized silicon layer at temperatures
of 850 $^{\circ}$C \cite{Velichko-10} and 800 $^{\circ}$C
\cite{Velichko-Hundorina-10,Velichko-Hundorina-11} with annealing
duration of 60 s. It is interesting to note that due to the higher
temperatures and longer duration a significant increase in the
thermal budget occurred in the experiments that were simulated in
\cite{Velichko-10} and
\cite{Velichko-Hundorina-10,Velichko-Hundorina-11}. Therefore, it
is reasonable to investigate the mechanism of ``tail'' formation
for low-energy boron implantation in crystalline silicon during
the subsequent low-temperature annealing with durations longer
than 10 s, i. e., for the increased thermal budget. It is worth
noting that the elucidation of the mechanisms of ``tail''
formation has a practical importance because this phenomenon
influences the depth of \textit{\textbf{p-n}} junction, as well as
it is actual needed for the investigation of the transport
processes of implanted boron atoms. Indeed, it has been shown in
\cite{Velichko-Hundorina-10,Velichko-Hundorina-11} that the
mechanism of boron diffusion varies with a change in the annealing
temperature.

\section{Model of boron diffusion}

For simulation of interstitial diffusion of boron atoms, we use
the model proposed in \cite{Velichko-10,Velichko-VITT-11}, which
includes the following system of equations:

\textbf{1) the conservation law for immobile impurity atoms}

\begin{equation}
\label{ConsLaw}
{\frac{{\partial C^{T}(x,t)}}{{\partial t}}} =
{\frac{{C^{AI}(x,t)}}{{\tau ^{AI}}}} - G^{AI}(x,t) \quad {\rm ,}
\end{equation}

\textbf{2) the diffusion equation for nonequilibrium boron
interstitials}

\begin{equation}
\label{DifEg} d^{AI}\,{\frac{{\partial ^{\, 2}C^{AI}}}{{\partial
x^{2}}}} - {\frac{{C^{AI}}}{{\tau ^{AI}}}} + G^{AI}(x,t) = 0 \quad
{\rm ,}
\end{equation}

\noindent which can be present in the form suitable for numerical
or analytical solution \cite{Velichko-2011},

\begin{equation}
\label{DifEgM} {\frac{{\partial ^{\, 2}C^{AI}}}{{\partial x^{2}}}}
- {\frac{{C^{AI}}}{{l_{AI}^{2}} }} + {\frac{{\tilde
{g}^{AI}(x,t)}}{{l_{AI}^{2}} }} = 0 \quad {\rm ,}
\end{equation}

\noindent where

\begin{equation}
\label{lAI} l_{AI} = \sqrt {d^{AI}\tau ^{AI}}  \, , \qquad
\tilde{g}^{AI}(x,t)= G^{AI}(x,t) \, \tau^{AI} \, .
\end{equation}

Here $C^{T}$ is the total concentration of substitutionally
dissolved impurity atoms, impurity atoms incorporated into
clusters and trapped by various defects including the extended
defects (immobile boron atoms); $C^{AI}$ is the summarized
concentration of nonequilibrium boron interstitials in different
charge states; $d^{AI}$ and $\tau^{AI}$ are the diffusivity and
average lifetime of nonequilibrium impurity interstitials,
respectively; $l_{AI}$ and $G^{AI}$  are the average migration
length and generation rate of interstitial impurity atoms,
respectively. The concentration of impurity interstitials (mobile
species) is also included in $C^{T}$.

We use the stationary diffusion equation for interstitial impurity
atoms in view of their large average migration length  $l_{AI}$
($l_{AI}\gg l_{fall}$) and due to the small average lifetime of
these nonequilibrium interstitials $\tau^{AI}$ ($\tau^{AI}\ll
\tau_{pr}$). Here $l_{fall}$ is the characteristic length relevant
to the significant decrease of impurity concentration in the
high-concentration region and $\tau_{pr}$ is the duration of
thermal treatment. It is also supposed that the function
$G^{AI}(x,t)$ that describes the generation rate of interstitial
impurity atoms changes slowly with time, increasing as compared to
the value of $\tau^{AI}$.

It is worth noting that according to the experimental data
\cite{Stolk-97,Hofker-75}, a significant part of boron atoms in
the ``tail'' region is electrically inactive, i.e., impurity atoms
do not occupy the substitutional position. A few possible
explanations can be given to this phenomenon. For example, the
average lifetime of the nonequilibrium boron interstitials
$\tau^{AI}\sim \tau_{pr}$, i.e., it is equal to tens of  minutes
at temperatures of 800 $^{\circ}$C \cite{Stolk-97}  and 900
$^{\circ}$C \cite{Hofker-75} . However, this assumption
contradicts the common point of view that the interstitial
impurity atoms are highly mobile Really, $l_{AI}$ is not too large
($l_{AI}\sim$20 nm), whereas the duration of annealing is longer
than enough and for the high value of diffusivity of boron
interstitials a significant value of $l_{AI}$ should be expected.
In our opinion, the assumption that the boron interstitials in the
``tail'' region are being incorporated with the defects of the
crystal lattice and, therefore, become immobile is more
reasonable. However, this problem requires further investigation.

In contrast to the   models of
\cite{Velichko-10,Velichko-VITT-11}, the Pearson type IV
distribution [20] is used to describe the spatial distribution of
impurity atoms after implantation:

\begin{equation}
\label{Pearson} C_{0}(x)=C(x,0)= C_{m}F^{P}(x,R_{p},\triangle
R_{p},Sk) \, ,
\end{equation}

\noindent  where

\begin{equation}\label{Maximum_Concentration}
C_{m}=\displaystyle \frac{Q}{\sqrt{2\pi}\Delta R_{p}}\times
10^{-8} \quad [\mu m^{-3}] \, .
\end{equation}

Here $F^{P}(x,R_{p},\triangle R_{p},Sk)$ is the function that
describes the Pearson type IV distribution and which is tabulated
in \cite{BKKT-85}; $Q$ is the dose of ion implantation
[ion/cm$^{2}$]; $R_{p}$ and $\triangle R_{p}$ are the average
projective range of implanted ions and straggling of the
projective range, respectively; $Sk$ is the skewness of the
Pearson type IV distribution

On the other hand, to simplify the description for the spatial
distribution of generation rate of nonequilibrium boron
interstitials we used the Gaussian distribution:

\begin{equation}
\label{Interstitial_Generation}
G^{AI}(x,t)= G^{AI}_{m}(t)\exp
\left[ -\frac{(x-R_{m})^{2}}{2\triangle R_{p}^{ \,2}}\right] \, ,
\end{equation}

\noindent shifted to the position of $R_{m}$ to adjust with the
maximum of distribution of implanted ions. Here $R_{m}$ is the
position of a maximum of the Pearson type IV distribution for
impurity concentration; $G^{AI}_{m}(t)$ is the value of the
generation rate for interstitial impurity atoms in the maximum of
the Pearson type IV distribution. As well as in the models of
\cite{Velichko-10,Velichko-VITT-11}, it is supposed that boron
atoms occur in the interstitial positions either directly during
implantation, or during rearrangement or annealing of the clusters
that incorporate impurity atoms.

\section{Analytical solution of diffusion equation}

It can be seen from expression (\ref{Interstitial_Generation})
that the generation rate of nonequilibrium boron interstitials is
approximated by the product of two multipliers, one of them
describes the dependence of the generation rate on time, whereas
the other describes the spatial distribution of the generation
rate of nonequilibrium boron interstitials:

\begin{equation}
\label{Int_Generation} G^{AI}(x,t)= G^{AI}_{m}(t)\, G^{AI}_{sp}(x)
\, ,
\end{equation}

Let us substitute   expression (\ref{Int_Generation}) into
Eq.(\ref{DifEg}) and integrate the resulting equation with respect
to time over\textbf{} the interval [t$_{{\rm 0}}$,t$_{{\rm k}}$]:

\begin{equation}
\label{DifEgIntegral} d^{AI}\,{\frac{{d^{\,
2}C^{AI}_{eff}}}{{dx^{2}}}}\left( {t_{k} - t_{0}} \right) -
{\frac{{C^{AI}_{eff}}}{{\tau ^{AI}}}}\left( {t_{k} - t_{0}}
\right) + G_{sp}^{AI} (x){\int\limits_{t_{0}} ^{t_{k}} {G_{m}^{AI}
(t)\,}} dt\; = 0 \, {\rm ,}
\end{equation}

\noindent or

\begin{equation}
\label{DifEgConst} d^{AI}\,{\frac{{d^{\,
2}C^{AI}_{eff}}}{{dx^{2}}}} - {\frac{{C^{AI}_{eff}}}{{\tau
^{AI}}}} + g_{mk}^{AI} \,G_{sp}^{AI} (x) = 0 \, {\rm ,}
\end{equation}

\noindent where

\begin{equation}
\label{AverageIntCon} C^{AI}_{eff}(x) = {\frac{{1}}{{\left( {t_{k}
- t_{0}} \right)}}}\;\;{\int\limits_{t_{0}} ^{t_{k}}  {C^{AI}
(x,t)\,}} dt \, {\rm ,}
\end{equation}

\begin{equation}
\label{AverageGenRate} g_{mk}^{AI} = {\frac{{1}}{{\left( {t_{k} -
t_{0}} \right)}}}\;\;{\int\limits_{t_{0}} ^{t_{k}}  {G_{m}^{AI}
(t)\,}} dt \, {\rm .}
\end{equation}

Here $C^{AI}_{eff}(x)$ and $g_{mk}^{AI}$ are, respectively, the
concentration of impurity interstitials and the value of the
generation rate for boron interstitials in the maximum of space
distribution which are time-averaged on the interval [t$_{{\rm
0}}$,t$_{{\rm k}}$].

It follows from the transformations performed that for the case of
the generation rate for boron interstitials, representing the
product of time- and space dependent multipliers, the system of
Eqs.(\ref{ConsLaw}) and (\ref{DifEg}) is similar to the system of
equations (\ref{ConsLaw}) and (\ref{DifEgConst}) with the constant
coefficient $g_{mk}^{AI}$. It is worth noting that
Eq.(\ref{DifEgConst}) can be solved analytically. For example,
such solution was obtained in [18] for

\begin{equation}
\label{GenRateSpace} g_{mk}^{AI} \,G_{sp}^{AI} (x) = g_{mk}^{AI}
\,\exp {\left[ { - {\frac{{(x - R_{m} )^{2}}}{{2\Delta R_{p}^{2}}
}}} \right]} \, {\rm .}
\end{equation}

It is well known that the intensity of transient enhanced
diffusion decreases with time. Let us suppose that

\begin{equation}
\label{TimeDep} G_{m}^{AI} (t) = G_{m0}^{AI} \,\exp \left( { -
{\frac{{t - t_{0}} }{{\tau _{gen}} }}} \right) \, {\rm ,}
\end{equation}

\noindent where $\tau _{gen} $ is the relaxation time for the
generation of nonequilibrium impurity interstitials and
$G_{m0}^{AI} \,$ is the value of the generation rate in the
maximum of the Pearson type IV distribution for ${\rm t} = {\rm
t}_{{\rm 0}} $.

The parameters $G_{m0}^{AI} \,$ and $\tau _{gen}$ can be derived
if we obtain two values of $g_{mk}^{AI}$ for two thermal
treatments with different durations. Indeed, let us suppose for
simplicity that $t_{0}$ =0 and $g_{m1}^{AI}$ and $g_{m2}^{AI}$ are
the values of the generation rate for boron interstitials in the
maximum of space distribution time-averaged on the intervals
[0,t$_{{\rm 1}}$] and [0,t$_{{\rm 2}}$], respectively. Then, the
following system of two independent algebraic equations is valid:

\begin{equation}
\label{TimeDep1} g_{m1}^{AI} = G_{m0}^{AI} \,\tau _{gen} {\left[
{1 - \exp \left( { - {\frac{{t_{1}} }{{\tau _{gen}} }}} \right)}
\right]} \quad {\rm ,}
\end{equation}

\begin{equation}
\label{TimeDep2} g_{m2}^{AI} = G_{m0}^{AI} \,\tau _{gen} {\left[
{1 - \exp \left( { - {\frac{{t_{2}} }{{\tau _{gen}} }}} \right)}
\right]} \quad {\rm .}
\end{equation}

To obtain the solution of the system (\ref{TimeDep1}),
(\ref{TimeDep2}), let us specify the quantity A as $A =
{{g_{m1}^{AI} \,t_{1}}  \mathord{\left/ {\vphantom {{g_{m1}^{AI}
\,t_{1}} {\left( {g_{m2}^{AI} \,t_{2}}  \right)}}} \right.
\kern-\nulldelimiterspace} {\left( {g_{m2}^{AI} \,t_{2}}
\right)}}$. Then, the value of $\tau _{gen}$ can be obtained by
solving the nonlinear equation:

\begin{equation}
\label{Nonlinear} \exp \left( { - {\frac{{t_{1}} }{{\tau _{gen}}
}}} \right) - A\exp \left( { - {\frac{{t_{2}} }{{\tau _{gen}} }}}
\right) + \left( {A - 1} \right) = 0 \quad {\rm .}
\end{equation}

M\"{u}ller's iteration method of successive bisections and inverse
parabolic interpolation \cite{Kristiansen-63} was used in this
work to solve Eq.(\ref{Nonlinear}). Then, we obtained the value of
$G_{m0}^{AI} \,$ from Eq.(\ref{TimeDep1}) or Eq.(\ref{TimeDep2}).

\section{Simulation results}

\begin{figure}[!ht]
\centering {
\begin{minipage}[!ht]{12.0 cm}
{\includegraphics[scale=1.0]{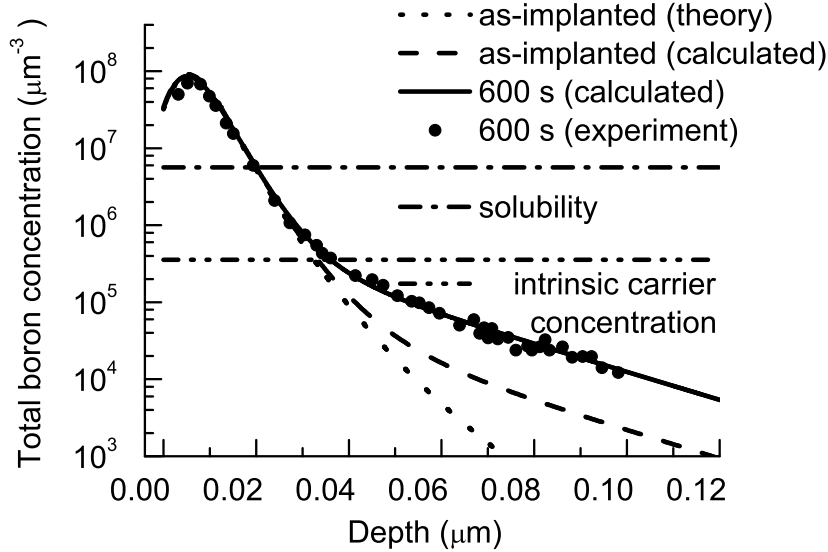}}
\end{minipage}} \caption{Calculated boron concentration profile after annealing
for 600 s at a temperature of 600 $^{\circ}$C. The dotted curve
represents boron concentration profile after ion implantation at
an energy of 1 keV with a dose of 1$\times$10$^{14}$ cm$^{-2}$
described by the Pearson type IV distribution. The dashed curve
represents boron concentration profile calculated under the
assumption of the long-range interstitial migration of a part of
the impurity atoms during ion implantation to provide fitting with
the experimental data for boron distribution after implantation
\cite{Napolitani-99}. The experimental data for boron
concentration after annealing (filled circles) are also taken from
\cite{Napolitani-99}}. \label{fig:B-600}
\end{figure}

\begin{figure}[!ht]
\centering {
\begin{minipage}[!ht]{12.0 cm}
{\includegraphics[scale=1.0]{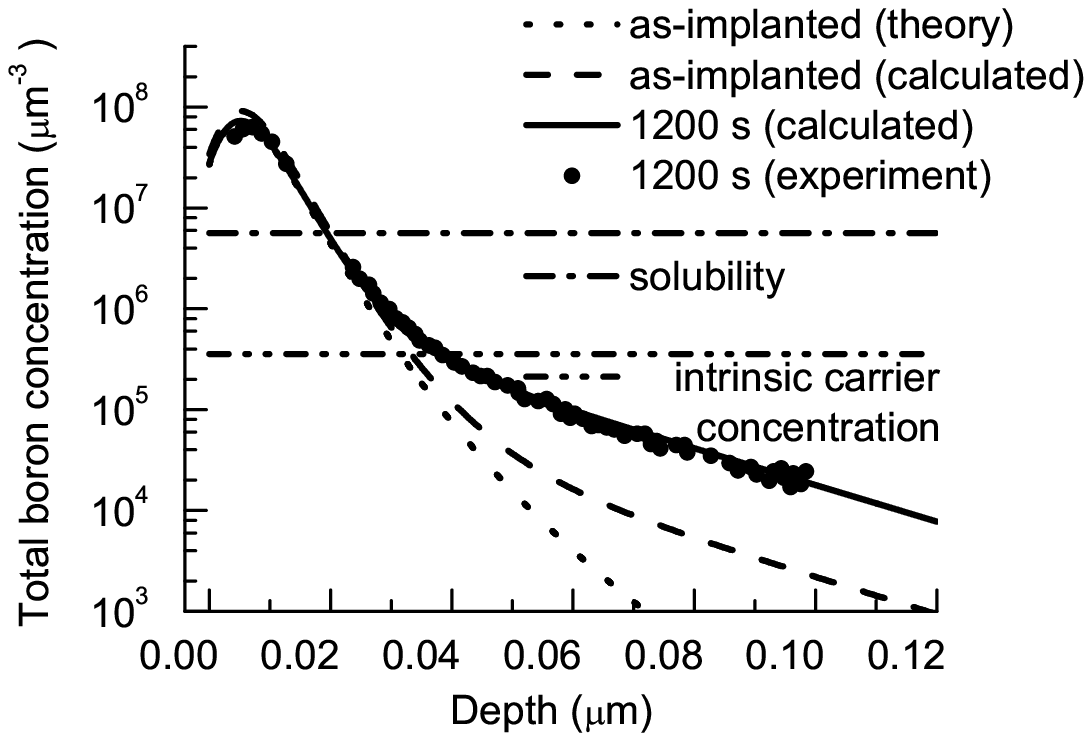}}
\end{minipage}} \caption{Calculated boron concentration profile after annealing
for 600 s at a temperature of 600 $^{\circ}$C. The dotted curve
represents boron concentration profile after ion implantation at
an energy of 1 keV with a dose of 1$\times$10$^{14}$ cm$^{-2}$
described by the Pearson type IV distribution. The dashed curve
represents boron concentration profile calculated under the
assumption of the long-range interstitial migration of a part of
the impurity atoms during ion implantation. The experimental data
for boron concentration after annealing (filled circles) are
taken from \cite{Napolitani-99}}. \label{fig:B-1200}
\end{figure}

The results of simulation of boron interstitial diffusion within
the framework of the model described above are presented in
Figs.~\ref{fig:B-600} and ~\ref{fig:B-1200}. For comparison we use
the experimental data of \cite{Napolitani-99}. In paper
\cite{Napolitani-99}, implantation of boron ions was carried out
with a dose of 1$\times$10$^{14}$cm$^{-2}$ and an energy of 1 keV
in silicon substrates of orientation (100) under $7^{\circ}$ tilt,
30$^{\circ}$ twist angles to prevent channeling. Before
implantation a pure silicon layer (10-$\mu$m thick,
\textbf{\textit{p}}-type conductivity with a resistivity of 10
$\Omega$m) has been grown up epitaxially. The chemical profiles of
the samples before and after annealing were measured by means of
the high depth resolution secondary ion mass spectrometry (SIMS).
The annealing of the samples was carried out at a temperature of
600 $^{\circ}$C for 10, 600, and 1200 s.

It follows from the experimental data of \cite{Napolitani-99} that
the boron concentration profile after implantation is
characterized by a significant asymmetry. For this reason, the
Pearson type IV distribution  was used to calculate the
concentration profile of ion-implanted boron. It was found in
\cite{Velichko-2012} that the values of $R_{p}$ = 0.007 $\mu$m,
$\triangle R_{p}$ = 0.0055 $\mu$m, and $Sk$ = 1.2 give the best
fit to the experimental data of \cite{Napolitani-99}. It is
necessary to note that the Pearson type IV distribution with
obtained values of the parameters gives a good fit for the whole
impurity concentration profile excluding the low-concentration
region. The problem of the accurate calculation of boron
concentration profile in the low-concentration region has been
also considered in \cite{Velichko-2012} where it was supposed that
heating of silicon substrate takes place during the ion
implantation described in \cite{Napolitani-99}. Then, due to the
elevated temperature, a fast interstitial diffusion of boron atoms
that occupy interstitial position occurs which results in the
formation of the ``tail'' in the low-concentration region of boron
profile. It was found in \cite{Velichko-2012} that the following
values of the parameters describing the process of interstitial
diffusion allow us to obtain a good fit to the boron profile in
the low-concentration region, namely: the average migration length
of boron interstitials $l_{AI}$  = 0.024 $\mu$m (24 nm) and a part
of boron atoms participating in the interstitial diffusion during
implantation $p_{AI}$ is approximately equal to 0.3 \%. In
addition, it was shown in \cite{Velichko-2012} that during
subsequent annealing at a temperature of 600 $^{\circ}$C for 10 s
the interstitial boron diffusion is characterized by the same
value of the average migration length of impurity interstitials
$l_{AI}$ = 24 nm. The best fit to the boron concentration profile
after annealing was obtained under the assumption that
approximately 0.7 \% of the implanted boron atoms entering into
the interstitial position in the high-concentration region of
ion-implanted impurity participated in the fast interstitial
diffusion during annealing and then became immobile again,
occupying substitutional positions or forming complexes with the
crystal lattice defects.

Taking into account the results of \cite{Velichko-2012}, in this
work we used the values of $R_{p}$ = 0.007 $\mu$m, $\triangle
R_{p}$ = 0.0055 $\mu$m, $Sk$ = 1.2, $l_{AI}$ = 0.024 $\mu$m, and
$p_{AI}$ = 0.308 \% to calculate the boron concentration profile
after implantation. For this value of skewness $Sk$, $R_{m}$ =
0.00521 $\mu$m. As can be seen from Figs.~\ref{fig:B-600} and
~\ref{fig:B-1200}, the results of simulation of boron diffusion
during annealing for 600 and 1200 s agree well with the
experimental data if the same value of the average migration
length of impurity interstitials $l_{AI}$ = 24 nm is used. Taking
into account the results of \cite{Velichko-2012} and the present
calculations, one can conclude that the migration length of
impurity interstitials varies negligibly with increase of
annealing duration. A good fit to the experimental data obtained
for different annealing durations with the same value of $l_{AI}$
confirms the correctness of the developed model of the long-range
migration for boron interstitial atoms in ion-implanted layers
during low-temperature annealing.

\section{Discussion}

It is worth noting that in paper \cite{Burenkov-80} the error
arising due to the use of Gaussian distribution instead of the
Pearson type IV distribution when modeling the redistribution of
ion-implanted impurity was investigated. It was supposed that the
Fick's law describes the diffusion of impurity atoms. As follows
from the results of \cite{Burenkov-80}, the error in calculating
the impurity concentration profile is less than 5\% if
$L>5|R_{m}-R_{p}|$. Here $L=\sqrt{D\, \tau_{pr}C^{AI}/C^{AI}_{i}}$
is the average length of the boron redistribution due to the pair
diffusion mechanism; $D$ is the diffusivity of impurity atoms;
$C^{AI}$ is the concentration of nonequilibrium silicon
self-interstitials; $C^{AI}_{i}$ is the thermally equilibrium
value of self-interstitials. Extrapolating the results of
\cite{Burenkov-80} to the case of diffusion of impurity
interstitials and taking into account that for the experimental
data of \cite{Napolitani-99} simulated in this paper
$|R_{m}-R_{p}|$ is equal to 1.79 nm whereas $l_{AI}$ = 24 nm, we
can conclude that using the Gaussian distribution to describe the
generation rate of boron interstitials is quite reasonable. In
addition, in our computations the Gaussian distribution of
generation rate is shifted to the position of $R_{m}$ and,
therefore, the error in calculating the concentration profile for
nonequilibrium boron interstitials is quite negligible.

\begin{figure}[!ht]
\centering {
\begin{minipage}[!ht]{12.0 cm}
{\includegraphics[scale=1.0]{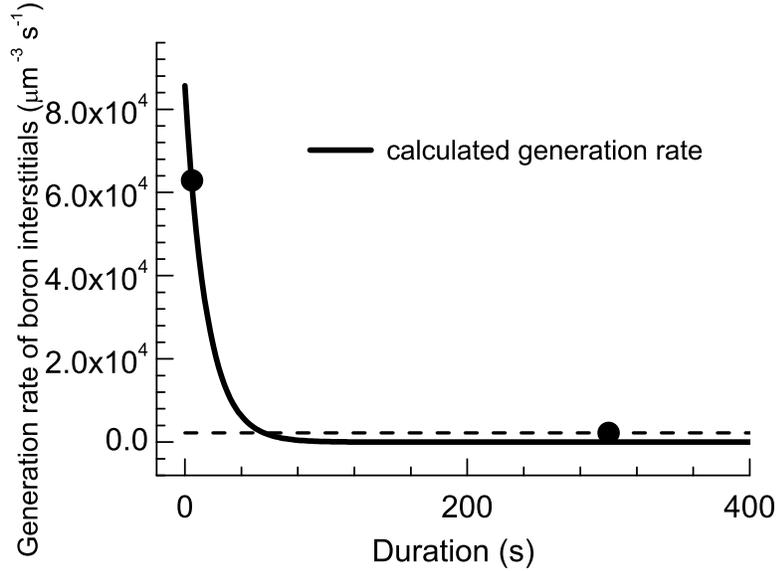}}
\end{minipage}} \caption{Calculated kinetics of the generation rate of
interstitial boron atoms during annealing at a temperature of 600
$^{\circ}$C. The filled circle represents the time-average
generation rate of boron interstitials for the interval from 0 s
to 10 s. The dashed curve with filled circle represents the
time-average generation rate of boron interstitials for the
interval from 0 s to 600 s}. \label{fig:B-TimeDep}
\end{figure}

The best fit to the boron concentration profiles after annealing
presented in Figs.~\ref{fig:B-600} and ~\ref{fig:B-1200} was
obtained under the assumption that approximately 1.5 \% and 2.29
\% of the implanted boron atoms participated temporally in the
fast interstitial diffusion during annealing for 600 and 1200 s,
respectively. It follows from these values and the value of
$p_{AI}$ = 0.7 \% for the part of boron atoms participating in
interstitial diffusion during annealing for 10 s
\cite{Velichko-2012} that there is a decrease in the generation
rate of boron interstitials with increase in annealing duration.
Indeed, the time-average interstitial generation rates for the
intervals [0,10] s, [0,600] s, and [0,1200] s are $g^{AI}_{m10}$ =
6.3$\times$10$^{5}$, $g^{AI}_{m600}$ = 2.19$\times$10$^{3}$, and
$g^{AI}_{m1200}$ = 1.675$\times$10$^{3}$ $\mu$m$^{-3}$s$^{-1}$,
respectively. The solution of the system (\ref{TimeDep1}),
(\ref{TimeDep2}) for the time intervals [0,10] s and [0,600] s
gives the following values of the parameters of time-dependent
function (\ref{TimeDep}) for the generation rate of boron
interstitials: $G^{AI}_{m0}$ = 8.5655$\times$10$^{4}$
$\mu$m$^{-3}$s$^{-1}$ and $\tau_{gen}$ = 15.33 s. The kinetics of
generation rate for nonequilibrium boron interstitials calculated
for these values of $G^{AI}_{m0}$ and $\tau_{gen}$ is presented in
Fig.~\ref{fig:B-TimeDep}. A similar calculation for the intervals
[0-10] s and [0-1200] s gives the following values of the kinetics
parameters: $G^{AI}_{m0}$ = 7.5473$\times$10$^{4}$
$\mu$m$^{-3}$s$^{-1}$ and $\tau_{gen}$ = 26.6 s. It follows from
the values obtained that the main fraction of boron interstitials
is generated during annealing duration shorter than 30 s.
Therefore, the additional experiments with annealing durations in
the range of 10 -- 600 s are needed to improve the accuracy of
kinetics determination.

It is worth noting that for simulation of boron redistribution
during annealing of ion-implanted layers the mechanism of
transient enhanced diffusion due to formation, migration and
dissociation of the ``boron atom -- silicon self-interstitial''
pairs is usually used. However, we can neglect this mechanism of
boron diffusion in simulating the experimental data of
\cite{Napolitani-99} due to the very low temperature of annealing
and, hence, due to the very small thermal budget. Indeed, the
boron diffusivity $D$ for a temperature of 600 $^{\circ}$C
calculated according to the results of \cite{Haddara-00} is equal
to 1.78$\times$10$^{-13}$ $\mu$m$^{2}$s$^{-1}$. Then the average
length of boron redistributions due to the pair diffusion
mechanism $L=\sqrt{D\, \tau_{pr}C^{AI}/C^{AI}_{i}}$  equals
1.4$\times$10$^{-3}$ $\mu$m (1.4 nm) for annealing duration of
1200 s and for concentration of nonequilibrium silicon
self-interstitials $C^{AI}$ is 10000 times higher than the
thermally equilibrium value of this species $C^{AI}_{i}$.
Therefore, it is possible to neglect the traditional transient
enhanced diffusion Really, one can see from Figs.~\ref{fig:B-600}
and ~\ref{fig:B-1200} that there is no boron redistribution in the
high-concentration region of impurity.

It is interesting to compare the derived value $l_{AI}$ = 24 nm
for B implanted in crystalline silicon with the values of the
average migration length of boron interstitials obtained for the
case of B implantation in the layer of silicon preamorphized by Ge
ions. For example, in \cite{Hamilton-07} $<$100$>$ oriented,
Czochralski grown silicon substrates of the
\textbf{\textit{n}}-type of conductivity and resistivity of about
10-25 $\Omega$m, were used for such experiments. These substrates
were implanted with Ge ions at 32 keV to a dose of 1$\times
$10$^{15}$ cm$^{-2}$ that provides amorphization to a depth of
$\sim$ 55 nm, as determined by Rutherford backscattering
spectrometry (RBS) and cross-sectional transmission electron
microscopy (XTEM). Boron was subsequently implanted at 500 eV to a
dose of 2$\times$10$^{15}$ cm$^{-2}$ at $^{\circ}$tilt and
$^{\circ}$twist. A 1.5 MeV He beam at 45$^{\circ}$ to the sample
were used for RBS. After implantation, samples received isochronal
rapid thermal annealing in the $\mathrm{N}_{2}$ ambient at a
temperature of 850 $^{\circ }$C for 60 s. Atomic profiles of B
were measured by SIMS. In \cite{Velichko-10} modeling of the
doping process described above was carried out, and the average
length of boron interstitials $l_{AI}$ = 12 nm was obtained. In
\cite{Yeong-08} boron implantation with an energy of 1 keV and a
dose of 1.5$\times$10$^{15}$ cm$^{-2}$ was carried out in (100)
oriented silicon substrates with the \textbf{\textit{n}}-type of
conductivity. For suppression of the transient enhanced diffusion
a near surface layer of silicon was preamorphized by implantation
of Ge ions with an energy of 15 êeV. After boron implantation, the
thermal annealing in the nitrogen ambient was carried out for 60 s
at a temperature of 800 $^{\circ }$C. Calculation of the
redistribution of ion-implanted boron for this experimental
environment performed in \cite{Velichko-12} has shown a good
agreement with the experimental data for the average migration
length of boron interstitials $l_{AI}$ = 11 nm.

It is clear from the presented results that the average migration
length of interstitial boron atoms in crystalline silicon is
significantly larger than in the silicon preamorphized by Ge ions
and recrystallized due to the solid phase epitaxy at the initial
stage of annealing. The obtained result confirms the strong
interaction of migrating boron interstitials with Ge atoms
previously implanted in the silicon layer. This interaction
provides approximately a quadruple reduction of the average
lifetime of boron interstitials and the relevant decrease in their
migration length.

\section{Conclusions}

Simulation of impurity redistribution for boron implanted in
crystalline silicon during annealing at a temperature of 600
$^{\circ}$C for 600 and 1200 s has been carried out. The results
obtained extend and deepen the investigation described in
\cite{Velichko-2012} where boron diffusion for 10-s annealing has
been simulated. The boron concentration profiles calculated within
the framework of the model of the long-range migration of impurity
interstitials agree well with the experimental data for
low-temperature thermal treatment of crystalline silicon. It is
interesting to note that the same value of the average migration
length of nonequilibrium boron interstitials $l_{AI}$ = 24 nm can
be used to explain, within the experimental error, the formation
of extended ``tails'' in the low-concentration region of boron
concentration profile for annealing duration of 10, 600, and 1200
s as well as for explaining the fast interstitial diffusion during
ion implantation It was also shown that approximately 1.5 \% and
2.29 \% of the implanted atoms occupied temporally the
interstitial position during annealing for 600 and 1200 s,
respectively, participated in the fast diffusion, and then became
immobile again, took up the substitutional position or formed
complexes with defects of the crystalline lattice. It follows from
the obtained values of the relaxation time for generation of
nonequilibrium impurity interstitials that the main fraction of
boron interstitials is generated during annealing duration shorter
than 30 s.

A comparison of the present calculations with the earlier
published data shows that the obtained value of $l_{AI}$ is
approximately two times greater than the average migration length
of boron interstitials in the silicon layers preamorphized due to
Ge ion implantation and recrystallized by means of solid phase
epitaxy at the initial stage of annealing. It means that the
migrating boron interstitials interact intensively with Ge atoms,
which have been used for the formation of a near-surface amorphous
layer before boron implantation.

\end{document}